\documentstyle[12pt,epsfig]{article}
\newcommand{\mh}{M_{H^{+}}}

\newcommand{\maz}{M_{A^0}}
\newcommand{\mHz}{M_{H^0}}
\newcommand{\mhz}{M_{h^0}}
\newcommand{\beq}{\begin{equation}}
\newcommand{\eeq}{\end{equation}}
\newcommand{\beqn}{\begin{eqnarray}}
\newcommand{\eeqn}{\end{eqnarray}}
\newcommand{\stackM}{\stackrel{\scriptstyle >}{{ }_{\sim}}}
\newcommand{\stackm}{\stackrel{\scriptstyle <}{{ }_{\sim}}}
\newcommand{\tb}{\tan\beta}
\newcommand{\ctb}{\cot\beta}
\newcommand{\ctbq}{\cot^2\beta}

\newcommand{\pl}{P_L}
\newcommand{\pr}{P_R}
\begin{document}

\thispagestyle{empty}
\def\pubnum{???}
\def\data{August, 1998}
\begin{flushright}
{\parbox{3.5cm}{
UAB-FT-450

KA-TP-14-1998

August, 1998

hep-ph/9808278
}}
\end{flushright}
\vspace{3cm}
\begin{center}
\begin{large}
\begin{bf}
TOP QUARK DECAY INTO CHARGED HIGGS BOSON IN A GENERAL
TWO-HIGGS-DOUBLET MODEL: IMPLICATIONS FOR THE TEVATRON
DATA\\
\end{bf}
\end{large}
\vspace{1cm}
J.A. COARASA, Jaume GUASCH , Joan SOL{\`A}\\
\vspace{0.25cm}
Grup de F{\'\i}sica Te{\`o}rica\\
and\\
Institut de F{\'\i}sica d'Altes Energies\\
\vspace{0.25cm}
Universitat Aut{\`o}noma de Barcelona\\
08193 Bellaterra (Barcelona), Catalonia, Spain\\
\vspace{1cm}
Wolfgang HOLLIK\\
Institut f{\"u}r Theoretische Physik,\\
 Universit{\"a}t Karlsruhe,
D-76128 Karlsruhe, Germany\\
\end{center}
\vspace{0.3cm}
\hyphenation{super-symme-tric de-pen-ding}
\hyphenation{com-pe-ti-ti-ve}
\begin{center}
{\bf ABSTRACT}
\end{center}
\begin{quotation}
\noindent
We analyze the  unconventional top quark decay mode $t \rightarrow H^+ b$ 
at the quantum level within
the  context of general Two-Higgs-Doublet models by including the full
electroweak effects from the Yukawa couplings.
The results are presented in the
on-shell renormalization  scheme with a physically well motivated definition of
$\tb$.  While the QCD corrections have been taken into account in the current
experimental analyses of
that decay, the electroweak effects have always been
neglected.  However, we find that they can be rather large and could
dramatically alter the interpretation of the present data 
from the Tevatron collider. For instance, in large portions of the parameter
space the electroweak effects prevent the Tevatron data from placing any
bound at all to the charged Higgs mass for essentially any value of $\tan\beta$.
\end{quotation}

\newpage

\baselineskip=6.5mm %(FOR PREPRINT)

With the discovery of the top quark at the Tevatron\cite{Tevatron} the 
last matter building block of the Standard Model (SM) has been fully accounted
for  by experiment.  Still, for an effective experimental underpinning of the
fundamental mass generation mechanism in the SM
one has to find the elementary Higgs scalar, which 
has intriguingly evaded all attempts up to now.
Therefore, in  spite of the great significance of the top quark discovery,
the Higgs mechanism -- the truly theoretical core of the SM -- 
remains
experimentally unconfirmed. On the other hand, the recent 
evidence for the possibility of neutrino oscillations 
\cite{neutrino} gives further support to the idea 
that the SM could be
 subsumed within a  larger and more fundamental theory. 
 The search for physics beyond the 
SM, therefore, has to continue with 
strong effort both at low and high energy.
And, complementary to the low-energy experiments,
the peculiar nature of the
top quark -- its large mass 
and its characteristic interactions with the scalar 
particles -- may help decisively to unearth further vestiges of physics 
beyond the SM.

\medskip 
The Two-Higgs-Doublet Model  ($2$HDM)\cite{Hunter} plays a special role 
as the simplest extension of the electroweak sector of 
the SM.
 In this 
class of models one enlarges the scalar sector of the SM by
 the introduction
of another  Higgs doublet, thus
rendering the  scalar sector in terms of
\beq
H_1=\left(\begin{array}{c}
H_1^0 \\ H_1^{-}
\end{array} \right)
\ \ \ (Y=-1)\,,\ \ \ \ \
H_2=\left(\begin{array}{c}
H_2^{+} \\ H_2^0
\end{array} \right)\ \ \ (Y=+1) \,\,.
\label{eq:H1H2}
\eeq
After spontaneous symmetry breaking one is left with two CP-even 
(scalar) Higgs bosons $h^0$, $H^0$, a CP-odd 
(pseudoscalar) Higgs boson $A^0$ and a pair of charged Higgs 
bosons $H^\pm$.  The parameters of these models consist of: i) the 
masses of the Higgs particles,
$M_{h^0}$, $M_{H^0}$, $M_{A^0}$ and $\mh$
(with the convention $M_{h^0}<M_{H^0}$), 
ii) the 
ratio of the two {vacuum expectation values}
\beq
\tb\equiv{\langle H_2^0\rangle \over 
\langle H_1^0\rangle}\equiv{v_2\over v_1}\,\,\,,
\eeq
and the mixing angle $\alpha$ between the two CP-even states. 
Two types of such models have been of special interest
\cite{Hunter} which  avoid potentially 
dangerous tree-level 
Flavour Changing Neutral Currents: In Type I $2$HDM only one of the 
Higgs doublets is coupled to the fermionic sector, 
whereas in Type II $2$HDM each Higgs doubled ($H_1$, $H_2$) is coupled 
to the up-type fermions and down-type fermions respectively, the 
Yukawa couplings being
\beq
\lambda_t\equiv {h_t\over g}={m_t\over \sqrt{2}\,M_W\,\sin{\beta}}\;\;\;\;\;,
\;\;\;\;\; \lambda_b^{\{{\rm I,\,II}\}}\equiv {h_b\over g}={m_b\over \sqrt{2}
\,M_W\,\{\sin{\beta},\cos{\beta}\}}\,\,.
\label{eq:Yukawas}
\eeq
Type II models do appear in specific 
extensions of the SM, 
such as the Minimal Supersymmetric Standard Model (MSSM) which is
currently under intensive study both theoretically and experimentally.

\medskip 
In case that 
the charged Higgs boson is light enough, the top
quark  could decay via the non-standard channel $t\rightarrow H^+\, b$.  Based 
on this possibility the CDF collaboration at the Tevatron has undertaken 
an experimental program which at the moment has been used to put limits on 
the parameter space of Type II models\,\cite{CDF}. 
The bounds are 
obtained by searching for an excess of the cross-section
$\sigma(p\bar{p}\rightarrow t \bar{t} X\rightarrow \tau \nu_{\tau} X)$ with
respect to
$\sigma(p\bar{p}\rightarrow t\bar{t}X\rightarrow 
l\nu_{l}\,X)$ ($l= e,\,\mu$).
The absence of such an excess determines an upper bound on 
$\Gamma(t\rightarrow H^+\,b\rightarrow\tau^+\,\nu_\tau\,b)$ 
and a corresponding 
excluded region of the parameter space $(\tb,\mh)$.
However, it has  been shown that the one-loop quantum corrections to that 
decay width can be rather large. This applies not only to the 
conventional QCD one-loop corrections\cite{CD}
-- the only ones used in Ref.\,\cite{CDF} -- 
but also to the QCD and electroweak corrections in the framework of the
MSSM\cite{GJS,CGGJS}. Thus the CDF limits could be substantially modified by
radiative  corrections\cite{GuaschSola} and in some cases the bound 
even disappears. 

To our knowledge, and in spite of some existing approximate 
calculations\,\footnote{See \,\cite{Yang} and
references therein. Because of the approximations used, 
neither of these references was really sensitive to the  potentially large
quantum effects reported here at low and high $\tan\beta$.},
a fully-fledged account of the main
electroweak corrections to $\Gamma(t\rightarrow H^+\,b)$
in the framework of general $2$HDM's 
is lacking in the literature.
Thus, in this letter we address the complete computation of the one-loop
electroweak contributions (EW) at leading order in both $\lambda_t$ and
$\lambda_b$ in generic Type I and Type II $2$HDM's and explore their impact on the
Tevatron data. Clearly, a detailed  treatment of $\Gamma(H^+ \rightarrow \tau^+\,
\nu_\tau)$ at the quantum level is also mandatory
to  perform this analysis in a consistent way.

We remark that although CLEO data on $BR(b\rightarrow s \gamma)$ could
preclude the existence of a light charged Higgs 
boson~\cite{CLEO} -- thus barring the possibility of the top quark
 decaying into it -- this assertion is not completely general and, moreover, 
needs
further experimental confirmation. In fact, there is no direct
experiment (at the level of the Tevatron analysis under consideration) 
supporting
the indirect implications on charged higgses from radiative B-meson decays.
Originally, the bounds from CLEO data were based on the computation up to 
leading
order (LO)  of $BR(b\rightarrow s \gamma)$\cite{bsgammaLO}. However, 
to this order
the theoretical result suffered from very  large
uncertainties~\cite{uncertainties}.  Recently the next to leading  order (NLO)
calculation has become  available~\cite{bsgammaNLO2HDM,BorzumatiBSG}, and the
theoretical situation 
seems to be settling.  The NLO calculation shows that Type I charged Higgs 
bosons masses are not restricted by $b\rightarrow s\,\gamma$ decay data 
either because of falling inside the experimental band or because of being 
not reliable.  As for Type II charged Higgs bosons, a lower bound of 
$\sim 255\, GeV$, with an 
error of at most several tens of $GeV$, has been achieved using 
the conservative ($95\%$ C.L.) CLEO allowed band 
$BR(b\rightarrow s \gamma)=(1.0-4.2)\times 10^{-4}$~\cite{CLEO}.  
Nevertheless, as stated,
the experimental situation  is not completely settled. 
Recently CLEO has presented the preliminary new result
$BR(b\rightarrow s \gamma)=(3.15\pm 0.35 \pm 0.32 \pm 0.26 ) \times 10^{-4}$
which modifies the upper limit above to $4.5\times 10^{-4}$~\cite{ichep98} and,
therefore, it weakens the previous bound on the charged Higgs mass.
On the other hand the ALEPH result is 
$BR(b\rightarrow s \gamma)=(3.11 \pm 0.80 \pm 0.72 ) \times
10^{-4}$~\cite{aleph98} which implies
an upper limit ($90\%$ C.L.) of $4.9\times 10^{-4}$.
Although both results are fully compatible, the latter entails
a
lower bound on the charged Higgs mass of $\sim 150\, GeV$ at large
$\tb$\, \cite{BorzumatiBSG}--thus allowing $t\rightarrow H^{+}\,b$ 
also for Type II models.  
It is our aim to investigate, independent of and complementary to
the indirect constraints, 
the decay $t\rightarrow H^+\,b$ in general $2$HDM's (Types I and
II) by strictly taking into consideration the direct data from
Tevatron on equal footing as in Ref.\,\cite{CDF}. 
This study is complementary to
the supersymmetric one in Ref.\cite{GuaschSola} and it should be useful to 
distinguish the kind of quantum effects expected in general $2$HDM's as
compared to those foreseen  within the context of the MSSM.

\medskip
The one-loop Feynman diagrams contributing to the decay 
$t \rightarrow H^+ b$ under consideration can be seen in Ref.~\cite{CGGJS}:
Fig.\,3 (all diagrams), Fig.\,4 (diagrams $C_{b3}$, 
$C_{b4}$, $C_{t3}$, $C_{t4}$), Fig.\,5 (diagram $C_{H1}$) and Fig.\,6 (diagram 
$C_{M1}$) of that reference. It goes without saying that the calculation of
these diagrams in general $2$HDM's is different from that in
Ref.\cite{CGGJS}, 
and this is so even for the Type II case since some of the
Higgs boson Feynman rules for supersymmetric models\,\cite{Hunter} cannot
be borrowed without a careful adaptation of the 
couplings\,\footnote{We have generated a
fully consistent set. In part they can be found in \cite{HollikZP} and
references therein.}. The interaction Lagrangian describing the
$H\,t\,b$-vertex  in Type-$j$ $2$HDM $(j=I,II)$ is:
\beq
{\cal L}_{H t b}^{(j)}=\frac{g}{\sqrt{2}M_W} H^-\,
 \bar{b}\, \left[m_t\, \ctb\,\pr +
 m_b\, a_j\, \pl \right]\, t + {\rm h.c.}
\label{eq:interaction}
\eeq
where we have introduced the parameter $a_j$ with $a_I\equiv -\ctb$, 
$a_{II}\equiv +\tb$.  From the 
interaction Lagrangian~(\ref{eq:interaction}) it is patent that 
for Type I models the 
branching ratios $BR(t\rightarrow H^+\,b)$ and $BR(H^+\rightarrow 
\tau^+ \nu_\tau)$ are relevant only at low $\tb$,  
whereas for Type II models the former branching ratio can be important both at 
low and high $\tb$ and the latter is only significant at high 
values of $\tb$.

\medskip
The renormalization procedure 
required for the one-loop amplitude
closely follows that of
Ref.~\cite{CGGJS}.
For Type II models the one-loop 
counterterm and vertex structures are formally as in
Ref.~\cite{CGGJS}, whereas for Type I there are some differences.
Nonetheless the two types of $2$HDM's can be treated 
simultaneously within a
unified formalism as follows: 
The counterterm Lagrangian $\delta{\cal L}_{Hbt}^{(j)}$ for each
$2$HDM model $j=I,II$ reads 
\beq
\delta{\cal L}_{Hbt}^{(j)}={g\over\sqrt{2}\,M_W}\,H^-\,\bar{b}\left[
\delta C_R^{(j)}\ m_t\,\ctb\,\,P_R+
\delta C_L^{(j)}\ m_b\,a_j\,P_L\right]\,t
+{\rm h.c.}\,,
\label{eq:LtbH2}
\eeq
with
\beqn
\delta C_R^{(j)} &=& {\delta m_t\over m_t}-{\delta v\over v}
+\frac{1}{2}\,\delta Z_{H^+}+\frac{1}{2}\,\delta Z_L^b+\frac{1}{2}
\,\delta Z_R^t
-{\delta\tb\over\tb}+\delta Z_{HW}\,\tb\,,\nonumber\\
\delta C_L^{(j)} &=& {\delta m_b\over m_b}-{\delta v\over v}
+\frac{1}{2}\,\delta Z_{H^+}+\frac{1}{2}\,\delta Z_L^t+\frac{1}{2}
\,\delta Z_R^b
\mp{\delta\tb\over\tb}\,-\delta Z_{HW}\,\frac{1}{a_j}\,,
\label{eq:deltacgen}
\eeqn
where in the last expression the upper minus sign applies to Type I models 
and the lower plus sign to Type II -- hereafter we will adopt 
this convention.

The counterterm $\delta\tb/\tb$ is defined in such a way that
it absorbs the one-loop contribution to
 the decay 
width $\Gamma(H^+\rightarrow \tau^+ \nu_\tau)$, 
yielding
\beq
{\delta\tb\over \tb}
=\mp\left[{\delta v\over v}-\frac{1}{2}\delta Z_{H^\pm}
+ \delta Z_{HW}\frac{1}{a_j}+
\Delta_{\tau}^{(j)}\,\right]\,\,. 
\label{eq:deltabeta}
\eeq
The quantity
\beq
\Delta_{\tau}^{(j)}=-{\delta m_{\tau}\over m_{\tau}} -\frac{1}{2}\delta 
Z_L^{\nu_{\tau}}-\frac{1}{2}\delta Z_R^{\tau}-F_{\tau}^{(j)}\,, 
\label{eq:deltatau}
\eeq
contains the (finite) process-dependent part of the counterterm, where
$F_\tau$ 
comprises the complete set of one-particle-irreducible three-point functions
of the charged Higgs decay into $\tau^+\,\nu_\tau$.
Other equivalent 
definitions of this counterterm are possible, but 
this one, apart from having a clear physical meaning, automatically 
incorporates the corrections that would arise in $\Gamma(H^+\rightarrow 
\tau^+\,\nu_\tau)$.

Substituting (\ref{eq:deltabeta}) into~(\ref{eq:deltacgen}) one 
finally gets for the Type-$j$ model
\beqn
\delta C_R^{(j)} &=& 2 \delta_{j\,II}\left[
                    \frac{1}{2}\,\delta Z_{H^+}-{\delta v\over v}
                               \right]
          +\delta Z_{HW}\left[\tb\pm \frac{1}{a_j} \right] \nonumber\\
            & &    +{\delta m_t\over m_t}+
               \frac{1}{2}\,\delta Z_L^b+\frac{1}{2}\,\delta Z_R^t\pm
               \Delta_{\tau}^{(j)}\nonumber\\
\delta C_L^{(j)} &=& {\delta m_b\over m_b}
               +\frac{1}{2}\,\delta Z_L^t+\frac{1}{2}\,\delta Z_R^b+
               \Delta_{\tau}^{(j)}\,\,.
\label{eq:counterterms}
\eeqn
We immediately see that for Type I models the one-loop correction is 
free of ``universal'' contributions as could be expected from our 
definition of $\tb$ . 

\medskip 
The correction to the decay width in each $2$HDM 
can be written in the following way:
\beqn
\delta_{\rm 2HDM}^{(j)}&=& \frac{\Gamma^{(j)}(t\rightarrow
 H^+\,b)-\Gamma_0^{(j)}(t\rightarrow H^+\,b)}
{\Gamma_0^{(j)}(t\rightarrow
 H^+\,b)}\nonumber\\
&=& \frac{N_L^{(j)}}{D^{(j)}} [ 2\,{\rm Re}(\Lambda_L^{(j)})] 
+ \frac{N_R}{D^{(j)}} [ 2\,{\rm
 Re}(\Lambda_R^{(j)})]+\frac{N_{LR}^{(j)}}{D^{(j)}} [ 2\,{\rm
Re}(\Lambda_L^{(j)}+\Lambda_R^{(j)})]\,\,,
\label{eq:defdelta}
\eeqn
where the lowest-order width
in the on-shell $\alpha$-scheme 
is
\beq
\Gamma^{(j)}_0(t\rightarrow H^+\,b)=\frac{\alpha}{s_W^2}
 \frac{D^{(j)}}{16\,M_W^2\,m_t}
\lambda^{1/2}(1,\frac{m_b^2}{m_t^2},\frac{\mh^2}{m_t^2})\,\,,
 \label{eq:treelevel}
\eeq
with
\beqn
D^{(j)}&=&(m_t^2+m_b^2-\mh^2)(m_t^2\,\ctbq+m_b^2\,a_j^2)+
    4 m_t^2\,m_b^2\,a_j\,\ctb\nonumber\\
N_L^{(j)}&=&(m_t^2+m_b^2-\mh^2)m_b^2\,a_j^2\nonumber\\
N_R&=&(m_t^2+m_b^2-\mh^2)m_t^2\, \ctbq\nonumber\\
N_{LR}^{(j)}&=& 2\,m_t^2\,m_b^2\,a_j\,\ctb\,\,.
\eeqn
The corresponding correction in the $G_F$-scheme is\,\cite{CGGJS}: 
$\delta(G_F)=\delta-\Delta r$.

The renormalized one-loop vertices $\Lambda_{L,R}$ 
for each type of model are obtained
after adding up the
counterterms~(\ref{eq:counterterms}) to the one-loop form factors:
\beqn
\Lambda_L&=&\delta C_L + F_L\nonumber\\
\Lambda_R&=&\delta C_R + F_R\,\,\,.
\eeqn
We refrain from explicitly quoting the rather lengthy expressions for the
unrenormalized vertex functions $F_L$ and $F_R$ 
(and $F_{\tau}\equiv F^{\tau}_L$ above) for each $2$HDM;
the calculation and conventions follow 
those in Ref.\cite{CGGJS}. 

\medskip
In the numerical analysis presented in Figs.\,1-4 we have put several cuts on 
our set of inputs.  From the
study of the Bjorken process $e^+\,e^-\rightarrow Z\,h^0$ and the Higgs boson
pair production $e^+\,e^-\rightarrow h^0\,A^0$  one obtains
\footnote{See Ref.\,\cite{MariaK} for a review and references therein.}
\beq
M_{h^0}+M_{A^0}\stackM 90-110\,GeV\,,
\label{eq:MhMA}
\eeq
and hence it cannot yet be excluded
the possibility of a light neutral Higgs scalar, say below $50\,GeV$, 
in general $2$HDM's.
As for $\tan\beta$ we have restricted in principle to the segment
\beq
0.1\stackm\tb\stackm 60\,\,\,.
\eeq
For Type I models the limits are very weak while for Type II
the limit at the low $\tan\beta$ end is obtained from measurements of the
process $e^+\,e^-\rightarrow Z\rightarrow h^0/A^0\,\gamma$ for
$\tau^+\,\tau^-$, light quarks and $b$-quark decay channels\,\cite{MariaK}.
We adopt the same low $\tb$ limit for Type I models since in this region
the analysis should be similar. 
For the three Higgs bosons coupling we have imposed that they do not exceed 
the maximum unitarity level permitted for the SM three Higgs boson coupling, 
i.e. 
\beq
|\lambda_{HHH}|\stackm|\lambda_{HHH}^{SM}(m_H=
1\,TeV)|=g\frac{3}{2}\frac{(1\,TeV)^2}{2\,M_W}\,\,.
\label{eq:hhh}
\eeq
This condition restricts both the ranges of masses and 
of $\tan\beta$.
Moreover, we have imposed that the extra induced contributions to
the $\rho$ parameter are bounded by the 
current experimental limit\,\footnote{Notice that this condition restrains 
$\Delta r$ within the experimental range and {\it a fortiori} the 
corresponding corrections in the $G_F$-scheme. The bulk of
the EW effects are contained in the  non-universal corrections
predicted in the $\alpha$-scheme.} : 
\beq
|\Delta\rho|\leq0.003\,\,.
\label{eq:deltarho}
\eeq
With these restrictions, which are independent and truly
 effective in our
calculation, we limit our
numerical analysis within a wide region of parameter space where 
the correction
(\ref{eq:defdelta}) itself
 remains perturbative, except in those places
where for demonstrational 
purposes we explicitly exhibit a departure from this 
requirement.

\medskip
Before exploring the implications for the Tevatron analyses, we wish to show
the great sensitivity (through quantum effects) of the decay 
$t\rightarrow H^+\,b$ to the particular structure of 
the underlying $2$HDM.
Therefore, in the following we summarize our 
systematic scanning over the
parameter space of $2$HDM's;
in some cases, just to illustrate
maximum effects, we have stretched their ranges to the very limits defined by
conditions (\ref{eq:MhMA})-(\ref{eq:deltarho}).
In all cases we present our results in a significant region of the parameter 
space where the branching ratios $BR(t\rightarrow H^+\,b)$ and 
$BR(H^+\rightarrow \tau^+\,\nu_\tau)$ are expected to be sizeable. This
entails relatively light charged Higgs bosons ($\mh\stackm 150\,GeV$) and a low
(high)  value of $\tb$ for Type I (II) models. 

In Fig.\,1  we display the evolution of the 
correction~(\ref{eq:defdelta}) with $\tb$ for Types I and II $2$HDM's 
and for two sets of inputs A and B for each model. We separately
show the (leading) EW contribution, $\delta_{\rm EW}$, 
and the total correction, $\delta_{\rm Total}\equiv
\delta_{\rm EW}+\delta_{\rm QCD}$, which incorporates
the conventional QCD effects\,\cite{CD}. In this figure we have skipped the
interval $2\stackm\tan\beta\stackm 10$ where the branching ratio of 
$t\rightarrow H^+\,b$ is too small to be of phenomenological interest.
In the relevant $\tan\beta$
segments, that is below and above the uninteresting one,
we find that the pure EW contributions can be rather 
large, to wit: For Type I models, the positive effects can reach $\simeq 30\%$, 
while the negative contributions may increase `arbitrarily' --
thus effectively enhancing to a great extent the modest QCD corrections-- 
still in a region of parameter space respecting the restrictions
(\ref{eq:MhMA})-(\ref{eq:deltarho}); For Type II models, instead,
the EW effects can be very large, for both signs, in the high
$\tan\beta$ regime. In particular, the huge positive yields
could go into a complete ``screening'' of the QCD corrections. 

In Fig.\,2 (resp.\,3) we present the evolution of the corrections 
in Type I (resp.\,II) models
as a function of the other parameters, namely the tangent
of the CP-even mixing angle (a), the charged Higgs mass (b), 
the CP-odd scalar mass (c) and the CP-even scalar masses (d).
Inputs A and B for each model are as in Fig.\,1 whenever they are fixed. 
Notice the rapid oscillation, yet qualitatively different for both models, 
around
$\tan\alpha=0$. On the other hand, the fast evolution that can
be seen as a function of the masses is due to the fact that the 
Higgs self-couplings in the three-point functions are proportional 
to the splitting of the Higgs 
masses. In Fig.\,2d (also in Fig.\,3d) the range of the CP-even masses is
plotted until the condition $M_{h^0}<M_{H^0}$ is exhausted or there is a
breakdown of relations (\ref{eq:hhh}) and/or (\ref{eq:deltarho}).

\medskip
Next we turn to the discussion of the dramatic implications that the EW
effects may have for the decay
$t\rightarrow H^+\,b$ at the Tevatron.  The original analysis of the data
(based on the non-observation of any excess of $\tau$-events) and its
interpretation in terms of limits on the $2$HDM parameter space 
was performed in Ref.\,\cite{CDF} (for Type II models) without including 
the EW corrections. In these references an
exclusion plot is presented in the
$(\tb,\mh)$-plane after correcting for QCD effects only.
To demonstrate the potential impact of the EW loops
on these studies we follow the method of
Ref.\cite{GuaschSola}. Although the data used by the
Tevatron collaborations is based on inclusive $\tau$-lepton tagging
\,\cite{CDF}, it will suffice for illustrative purposes to 
concentrate ourselves on the ($\tau$,$l$)-channel\,\cite{GuaschSola,DPRoy}. 
In this way the comparison of the results for generic Type II $2$HDM's
and those already available for the specific case of the MSSM Higgs
sector\,\cite{GuaschSola} will be more transparent.
The production cross-section of 
the top quark in the  ($\tau$,$l$)-channel can be easily related to
the decay rate of $t\rightarrow H^+\,b$ and the branching ratio of 
$H^+\rightarrow \tau^+\,\nu_{\tau}$ as follows:
\beq
\sigma_{l\tau}=\left[\frac{4}{81}\,\epsilon_1+\frac{4}{9}\,
{\Gamma (t\rightarrow H^+\,b)\over
\Gamma (t\rightarrow W^+\,b)}\,
BR(H^+\rightarrow \tau^+ \,\nu_\tau)\,\epsilon_2\right]\,\sigma_{t\bar{t}}\,,
\label{eq:bfrac}
\eeq
This formula generalizes eq.(7) of Ref.\,\cite{GuaschSola} for the case
that $BR(H\rightarrow \tau \,\nu_\tau)$ is not $100\%$, as it 
indeed happens when
we explore the low $\tb$ region. In general we have
\beq
BR(H^+\rightarrow \tau^+ \,\nu_\tau)=
\frac{\Gamma(H^+\rightarrow \tau^+\,\nu_\tau)}
{\Gamma(H^+\rightarrow \tau^+\,\nu_\tau)+\Gamma(H^+\rightarrow
c\,\bar{s})}\,\,, \eeq
where we use the QCD-corrected amplitude for the last term in the 
denominator\,\cite{Gambino}.  This branching ratio is about $50\%$ for Type I
$2$HDM at low  $\tb$, and $100\%$ for Type II at high 
$\tb$ (the case studied in \cite{GuaschSola}).

\medskip
Finally in Figs.\,4a and 4b we have plotted the
perturbative exclusion regions in the
parameter space $(\tb,\mh)$ for intermediate and extreme sets of $2$HDM
inputs A, B, B' and C. In Type I models (a) we see that the bounds obtained from
the EW-corrected amplitude are generally less restrictive than those obtained by
means of tree-level and  QCD-corrected amplitudes. Evolution of the excluded
region from set A to set C in Fig.\,4a shows that the region tends to evanesce,
which is indeed the case when we further increase $M_{A^0}$ in set C. 
In Type II models (b) we also show a series of possible scenarios.
We have checked that the maximum positive effect $\delta_{EW}>0$ (set A
in Fig.4b) may completely cancel the QCD corrections and restore the full
one-loop width $\Gamma^{(II)}(t\rightarrow H^+\,b)$ to the tree-level
value (\ref{eq:treelevel}) just as if there were no QCD corrections 
at all!
Intermediate possibilities (set B') are also shown. In the other extreme
the (negative) effects $\delta_{EW}<0$ enforce the exclusion region
to draw back to curve C where it starts to gradually
disappear into a non-perturbative corner of the parameter space where one
cannot claim any bound whatsoever!!.  

\medskip
Some discussion may be necessary to compare the present analysis with the
supersymmetric one in Refs.\,\cite{CGGJS,GuaschSola}.
In the MSSM case, the Higgs
sector is of Type II. However, due to supersymmetric restrictions in the
structure of the Higgs potential, there are large cancellations
between the one-particle-irreducible vertex functions, so that the overall
contribution from the MSSM Higgs sector to the  correction (\ref{eq:defdelta})
is negligible. In fact, we have checked that when we take the Higgs 
boson masses
as they are correlated by the MSSM we recover the previous 
result\,\cite{CGGJS}.
Still, in the SUSY case there emerges a large effect
from the genuine sparticle sector, mainly from the SUSY-QCD contributions to 
the bottom mass renormalization counterterm\,\cite{CGGJS},  
which can be positive or negative because the correction flips sign with the
higgsino mixing parameter. In contrast, for general (non-SUSY) Type II models the
bulk of the EW correction comes from large unbalanced contributions from the
vertex functions, which can also 
flip sign with $\tan\alpha$ (Cf. Fig.\,3a) -- a free parameter
in the non-supersymmetric case. Although the size and sign of the effects can
be similar for a general Type II and a SUSY $2$HDM, they should be
distinguishable since the large corrections are attained for very different
values of the Higgs boson masses. For instance, in generic $2$HDM's (of both
types) large negative effects may occur for large
values of the CP-odd Higgs mass (Cf. Figs. 2c and 3c). In the MSSM the latter
should be essentially degenerate with the charged Higgs in that region and so
$t\rightarrow H^+\,b$ would have never occurred.  

Therefore, just to
illustrate one possibility, let us 
envision the following scenario. Suppose that
$t\rightarrow H^+\,b$ is not observed at the Tevatron -- 
or that it comes out highly suppressed beyond QCD expectations
 (Cf. Fig.4b, curve C) --  while at the same time
$H^+$ and $A^0$ are observed (maybe produced at the Tevatron itself or at LEP)
and both show up with a similar mass below $m_t$. Then these bosons
could well be supersymmetric Higgs bosons. If, however, a similar situation would
be encountered  but $A^0$ is not produced (because it is perhaps too heavy), then
the observed $H^+$ cannot probably be a SUSY Higgs. Notice (Cf. Fig. 4a) that
in this case the $H^+$ could also belong to a Type I model. In this case
further investigation would be required to disentangle the type of  non-SUSY
Higgs model at hand e.g. aiming at a determination of $\tan\beta$.  One
possibility would be from the decay $H^+\rightarrow \tau^+\,\nu_{\tau}$ after
$H^+$ been produced from mechanisms other than top quark decay; alternatively,
once a heavy $A^0$ would be found it would provide a handle to a
$\tan\beta$ measurement through the decays $A^0\rightarrow
\tau^+\,\tau^-$ and/or  $A^0\rightarrow b\,\bar{b}$. If these decays would be
tagged at a high rate, the non-SUSY model should necessarily be of Type II. 
At the LHC,  or at a NLC, one could even use $\Gamma(A^0\rightarrow
b\,\bar{b})/\Gamma(A^0\rightarrow t\,\bar{t}) \propto\tan^4\beta$ for very
heavy $A^0$.  In short, a combined procedure based on
quantum effects and direct production could be a suitable strategy to unravel
the identity of the Higgs bosons.

\medskip
To summarize, we have computed the electroweak one-loop corrections
to the unconventional top quark decay width 
$\Gamma(t\rightarrow H^+\, b)$ at the 
leading order in the Yukawa couplings both for general Type I and
Type II $2$HDM's.
We have found  that the EW corrections can be comparable in size to the
QCD effects and be of both signs.
The positive
ones can reach $30\%$ and $50\%$  for Type I and Type II models
respectively, which means that they could simply delete the QCD corrections.
The negative
ones can even be larger  ($-50\%$ or more in ample regions of
parameter space) for both models. We have also
shown that this fact may deeply influence the current interpretation
of the Tevatron data on that decay. Most important, we have argued that
knowledge of the EW quantum effects may be crucial to understand the nature
(supersymmetric or else) of
the Higgs bosons, if they are eventually found in future experiments at hadron
and/or $e^+\,e^-$ colliders.

\bigskip \noindent
{\bf Acknowledgements}:

\noindent 
The work of J.G. has been financed by a grant of the Comissionat 
per a Universitats i Recerca, Generalitat de Catalunya.  This work has also 
been partially supported by CICYT under project No.  AEN95-0882.

%%%%%%%%%%%%%%%%%%%%%%%%%%%%%%%%%%%%%%%%%%%%%%%%%%%%%%%%%%%%%%%%%%%

\vspace{0.75cm}
\begin{center}
\begin{Large}
{\bf Figure Captions}
\end{Large}
\end{center}

\begin{itemize}

\item{\bf Fig.\,1} The correction $\delta$, eq.\,(\ref{eq:defdelta}), 
  to the decay width 
  $\Gamma(t \rightarrow H^+ b)$ as a function of $\tan\beta$, 
  for Type I $2$HDM's (left hand side of the figure)
  and two sets of inputs $\{(\mh,\mHz,\mhz,\maz);\, \tan\alpha\}$, namely
  set A: $\{(70,175,100,50)\,GeV;\,3\}$ and
  set B: $\{(120,200,80,250)\,GeV; 1\}$.
  Similarly for Type II models (right hand side of the figure)
  and for two different sets of inputs,
  set A: $\{(120,300,50,225)\,GeV;\,1\}$ and set
  B: $\{(120,300,80,225)\,GeV;\,-3\}$.  
  Shown are the electroweak contribution $\delta_{\rm EW}$ and the 
  total correction $\delta_{\rm Total}=\delta_{\rm EW}+\delta_{\rm QCD}$.

\item{\bf Fig.\,2} The corrections  $\delta_{\rm EW}$ and 
$\delta_{\rm Total}$
  for the Type I $2$HDM as a function of 
  {\bf (a)} $\tan\alpha$,
  {\bf (b)} the charged Higgs mass, 
  {\bf (c)} the pseudoscalar Higgs mass, and 
  {\bf (d)} the heavy (labelled with a triangle) and light (unlabelled) scalar
  Higgs masses. 
  Inputs as in Fig.\,1 with $\tb=0.1$ for set A and $\tb=0.2$ for set B.

\item{\bf Fig.\,3} The corrections  $\delta_{\rm EW}$ and 
$\delta_{\rm Total}$
  for the Type II $2$HDM as a function of 
  {\bf (a)} $\tan\alpha$,
  {\bf (b)} the charged Higgs mass, 
  {\bf (c)} the pseudoscalar Higgs mass, and 
  {\bf (d)} the heavy (labelled with a triangle) and light (unlabelled) scalar
  Higgs masses. 
  Inputs as in Fig.\,1 with $\tb=35$ for both sets.

\item{\bf Fig.\,4} The $95\%$ C.L. exclusion plot in the
  $(\tb, \mh)$-plane for 
  {\bf (a)} Type I $2$HDM using three sets of inputs: 
  A and B as in Fig.\,1, and
  C: $\{(\mh,200,80,700)\, GeV;\,1\}$;
  {\bf (b)} Similarly for Type II models including three sets of inputs: 
  A as defined in Fig.\,1,
  B':$\{(\mh,200,80,150)\,GeV;\,0.3\}$, and 
  C:$\{(\mh,200,80,150)\,GeV;\,-3\}$.  
  Shown are the tree-level, QCD-corrected and fully 2HDM-corrected 
  contour lines.  The excluded region in each case 
  is the one lying below these curves.

\end{itemize}

\newpage

\pagestyle{empty}
\begin{center}

\epsfig{file=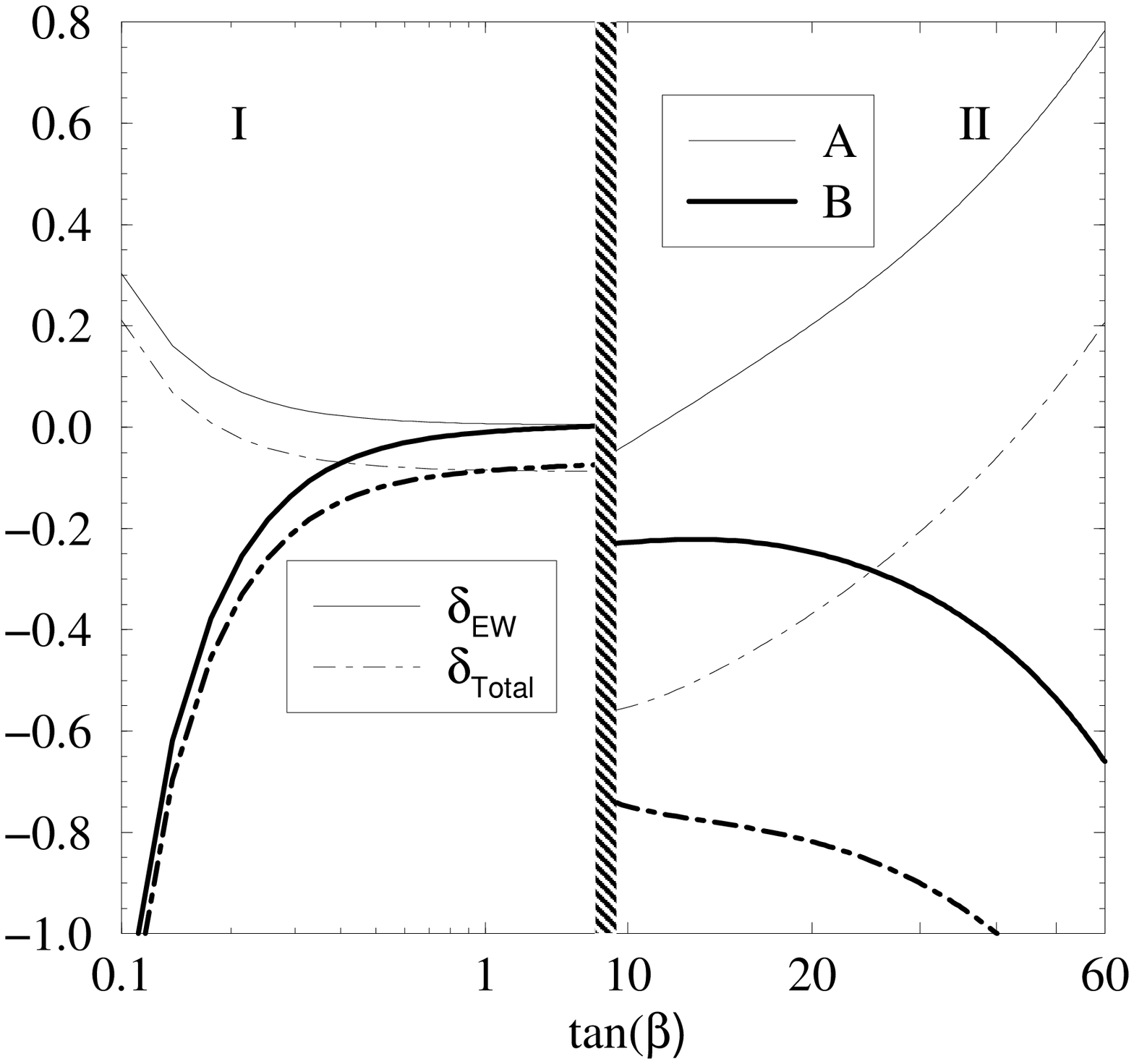,width=13cm}

{\Large Fig. 1}

\newpage

\begin{tabular}{c}
\epsfig{file=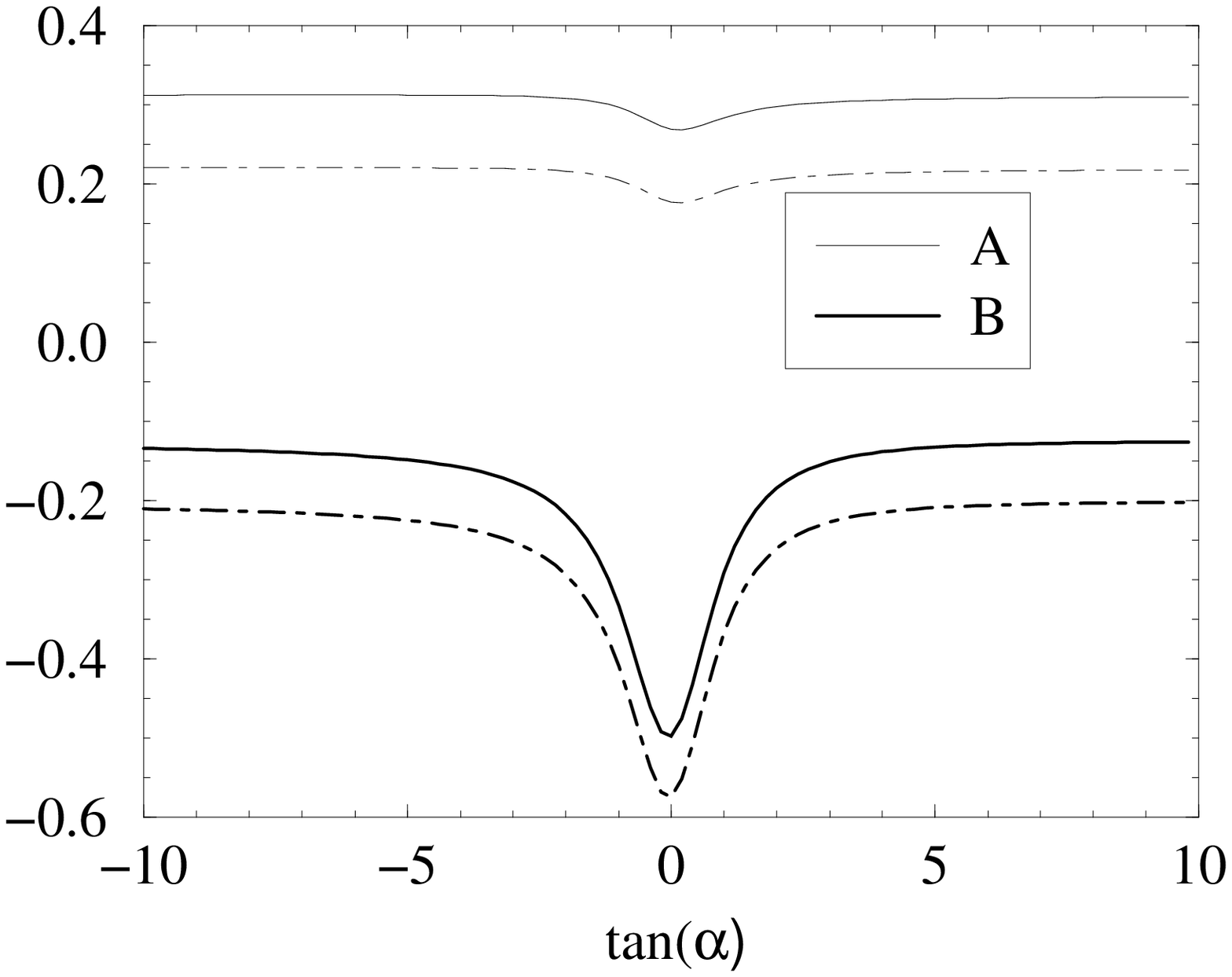,width=9.5cm} \\{\large (a)} \\ \vspace{.2cm}\\
 \epsfig{file=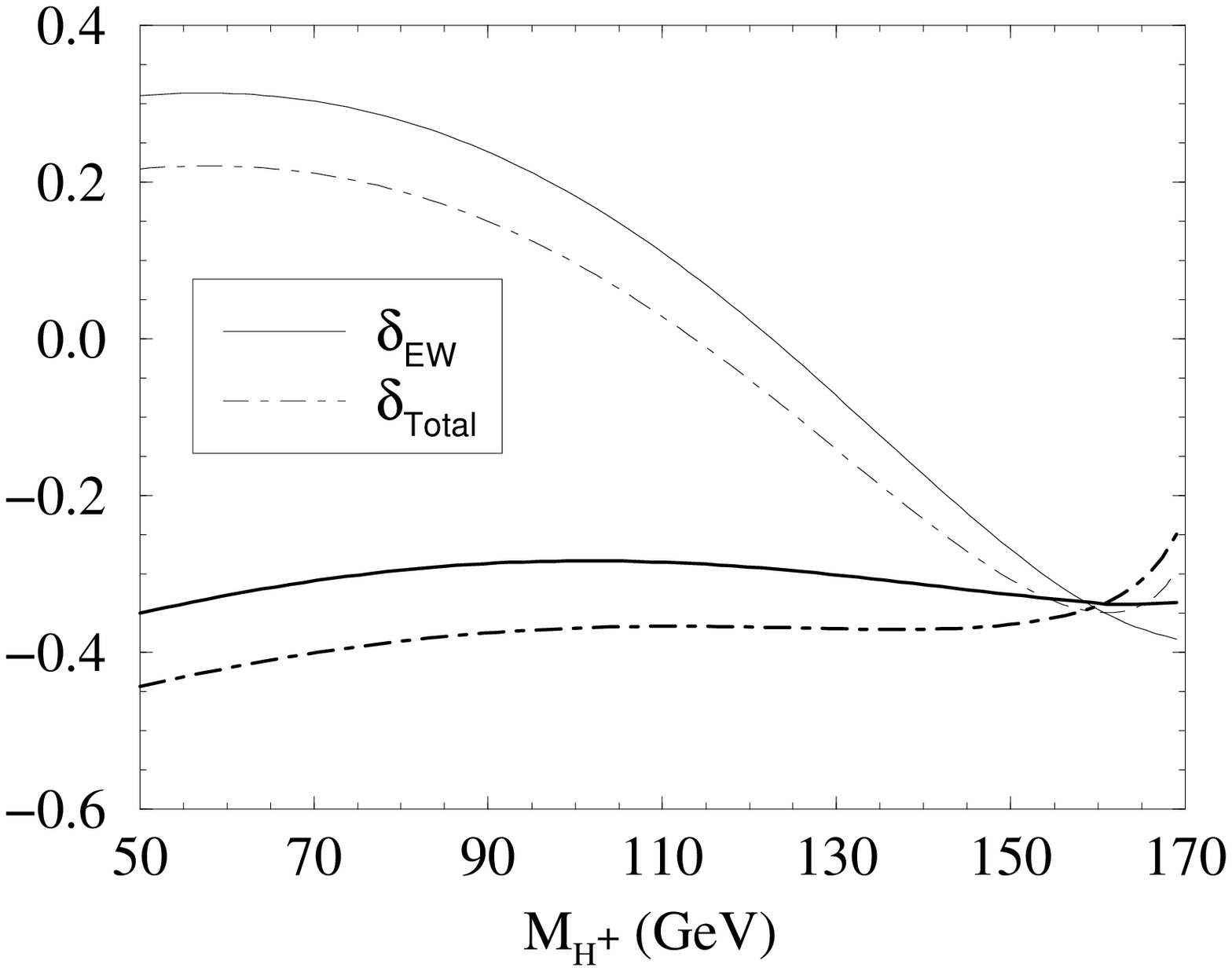,width=9.5cm} \\ {\large (b)}\\  \vspace{.2cm}\\
{\Large Fig. 2}
\end{tabular}

\newpage

\begin{tabular}{cc}
\epsfig{file=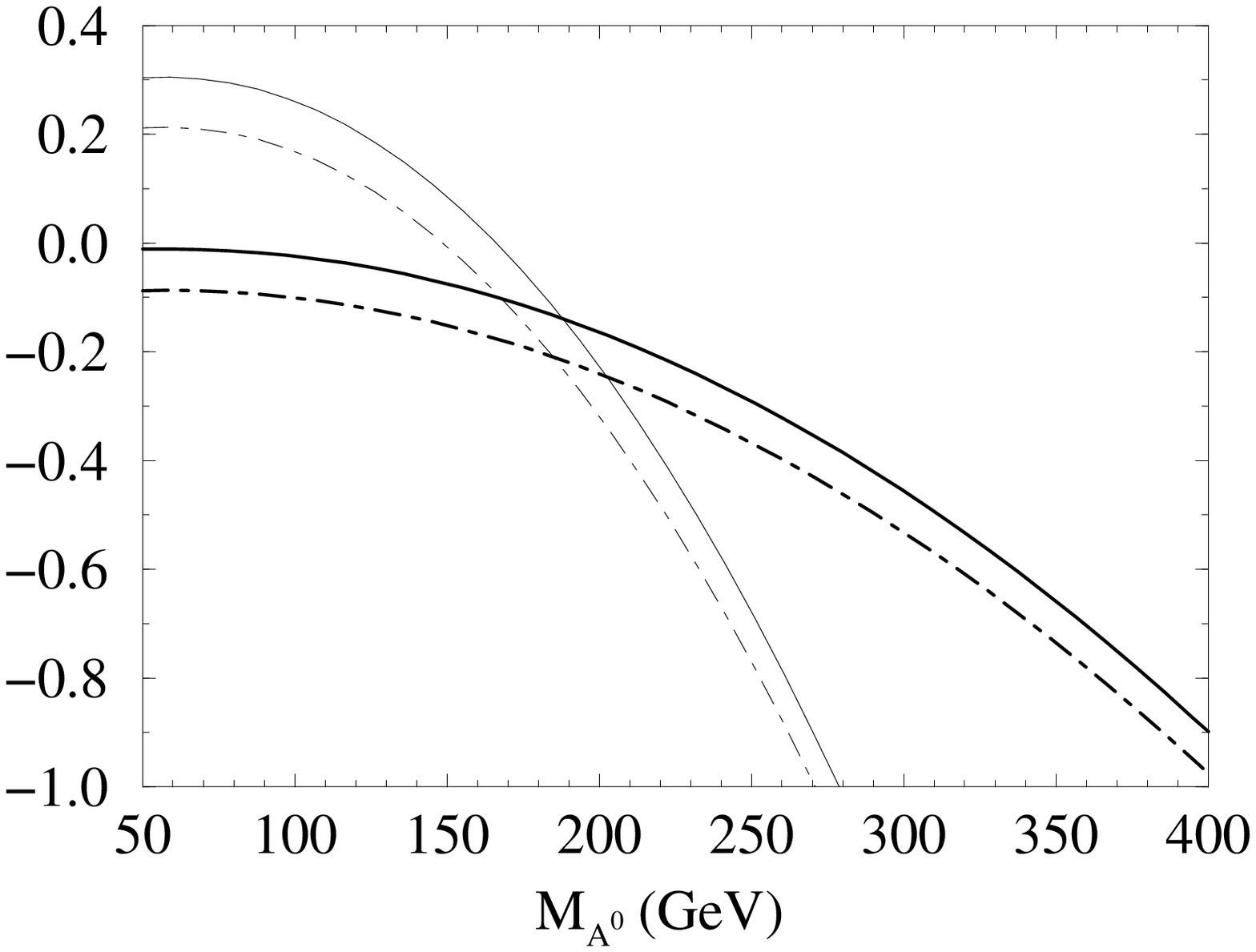,width=9.5cm}\\ {\large (c)}\\ \vspace{.2cm}\\
 \epsfig{file=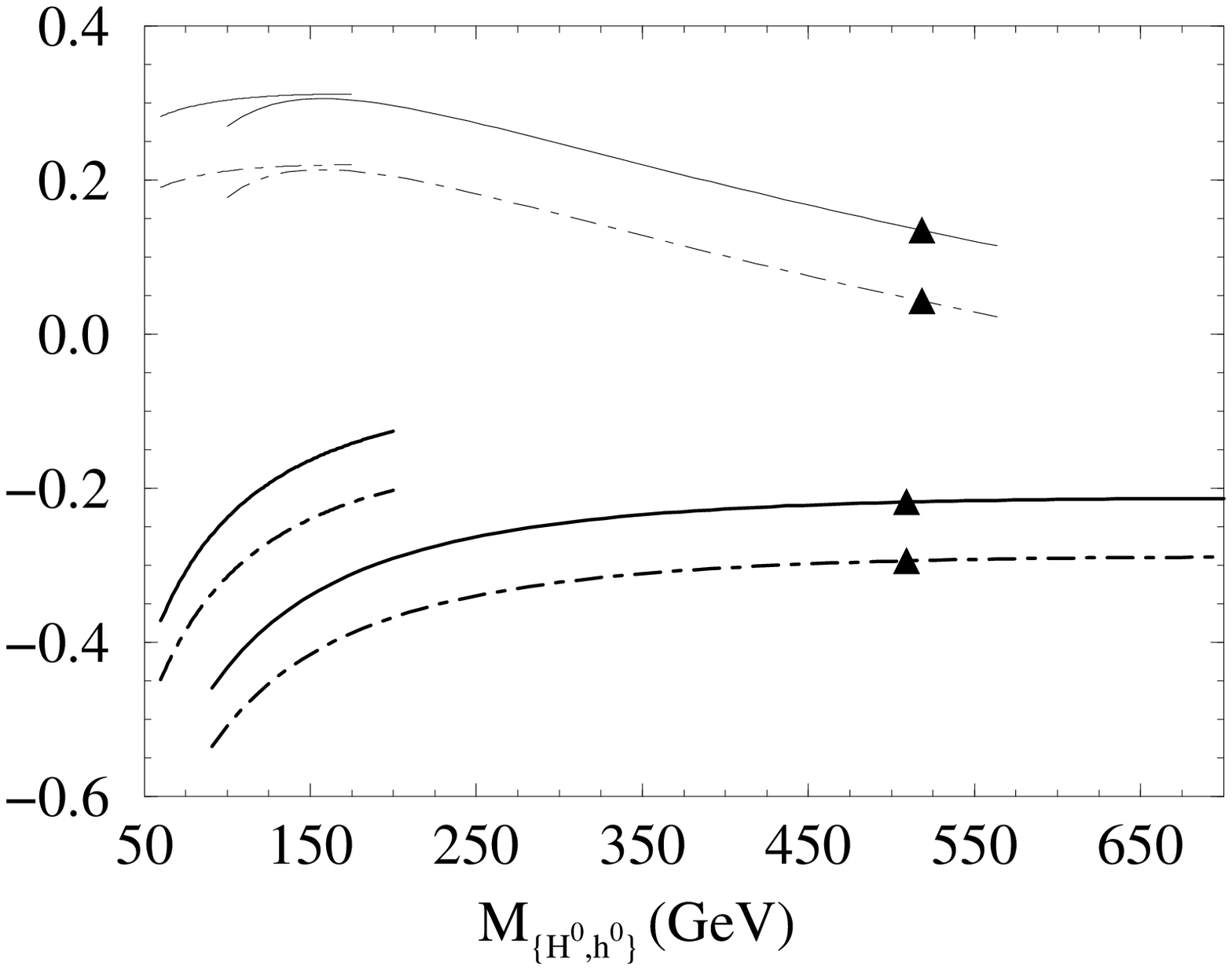,width=9.5cm} \\ {\large (d)}\\ \vspace{.2cm}\\
{\Large Fig. 2 (cont.)}
\end{tabular}

\newpage

\begin{tabular}{cc}
\epsfig{file=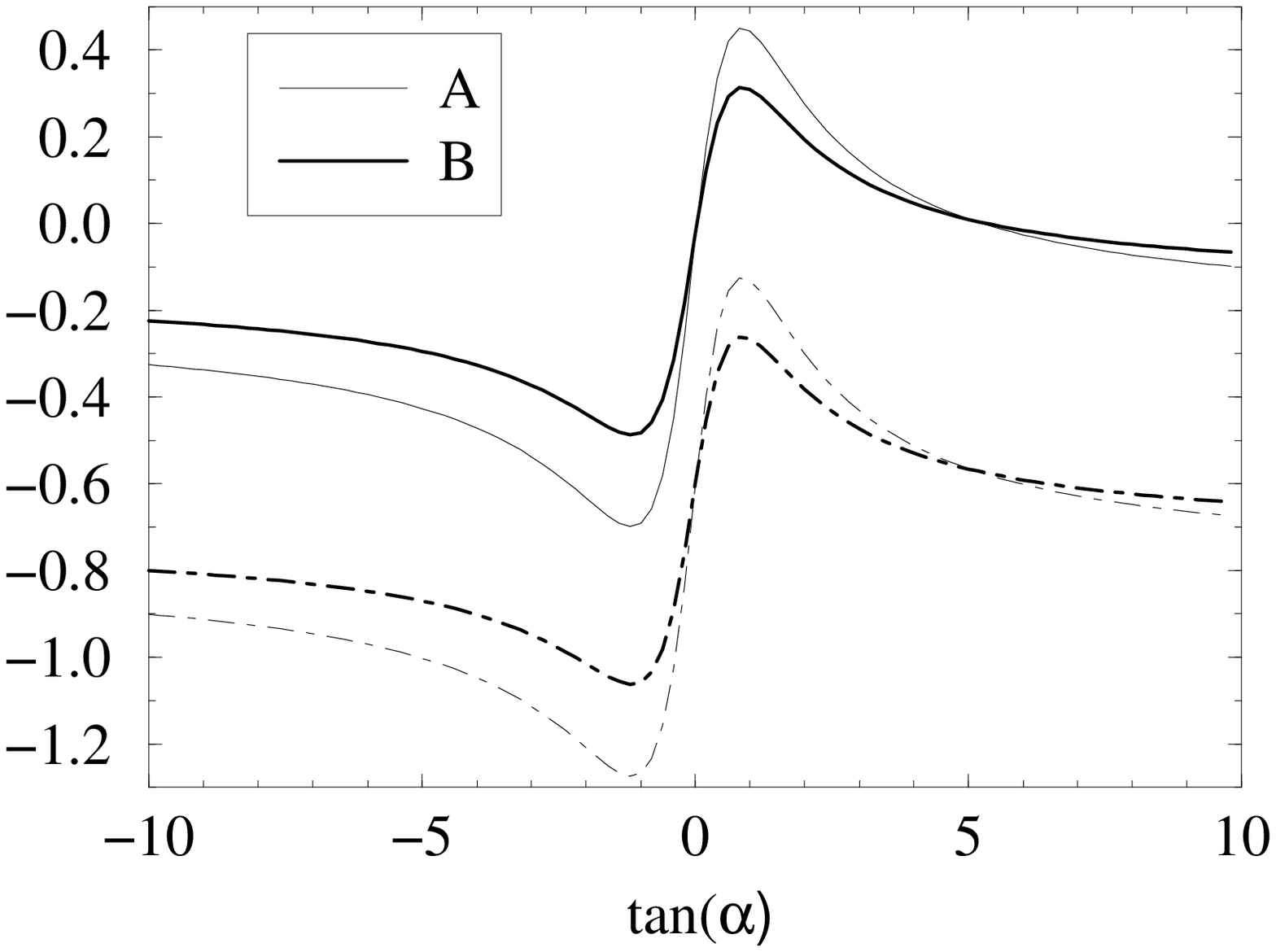,width=9.5cm} \\{\large (a)} \\\vspace{.2cm}\\
 \epsfig{file=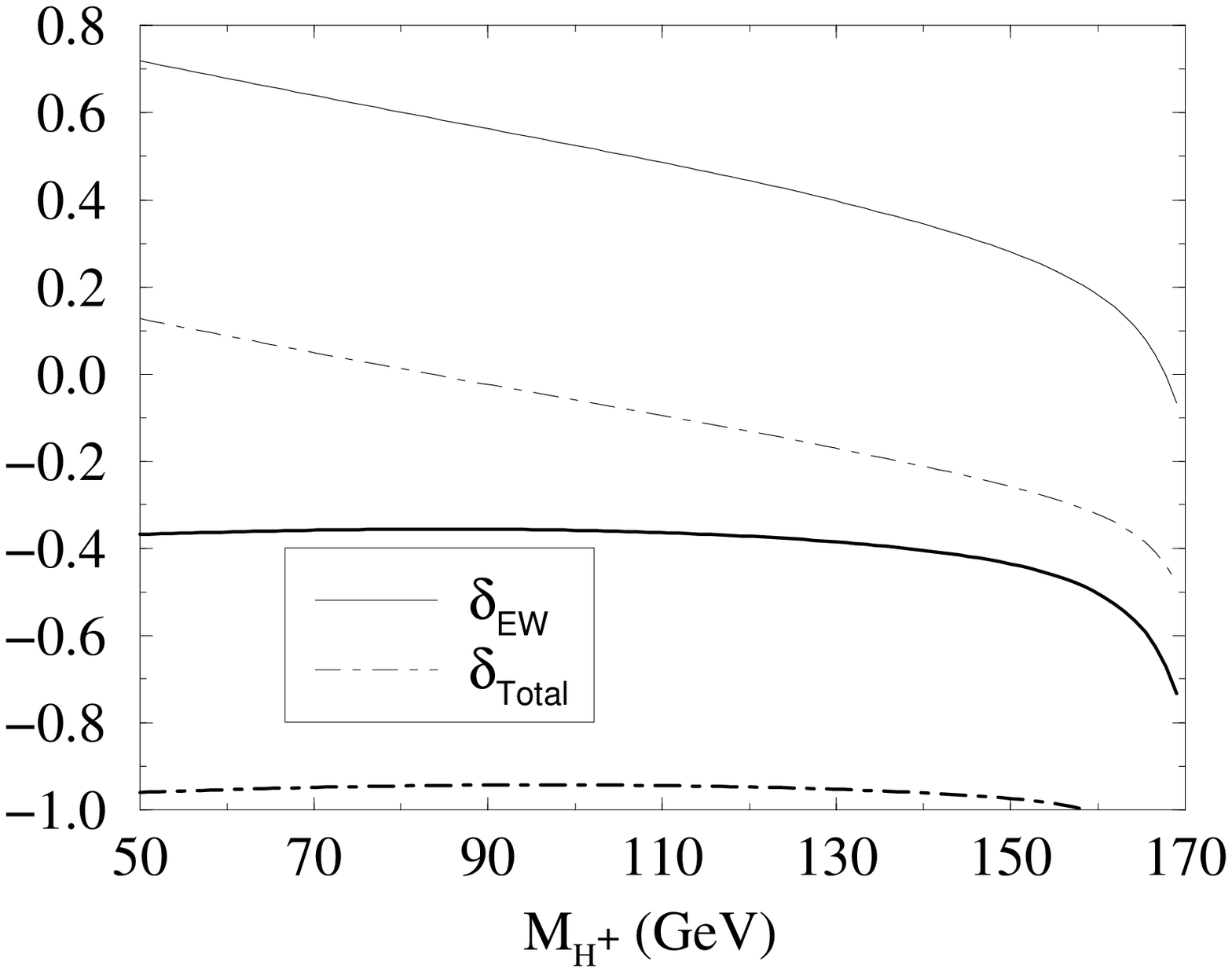,width=9.5cm} \\ {\large (b)} \\\vspace{.2cm}\\
{\Large Fig. 3}
\end{tabular}

\newpage

\begin{tabular}{cc}
\epsfig{file=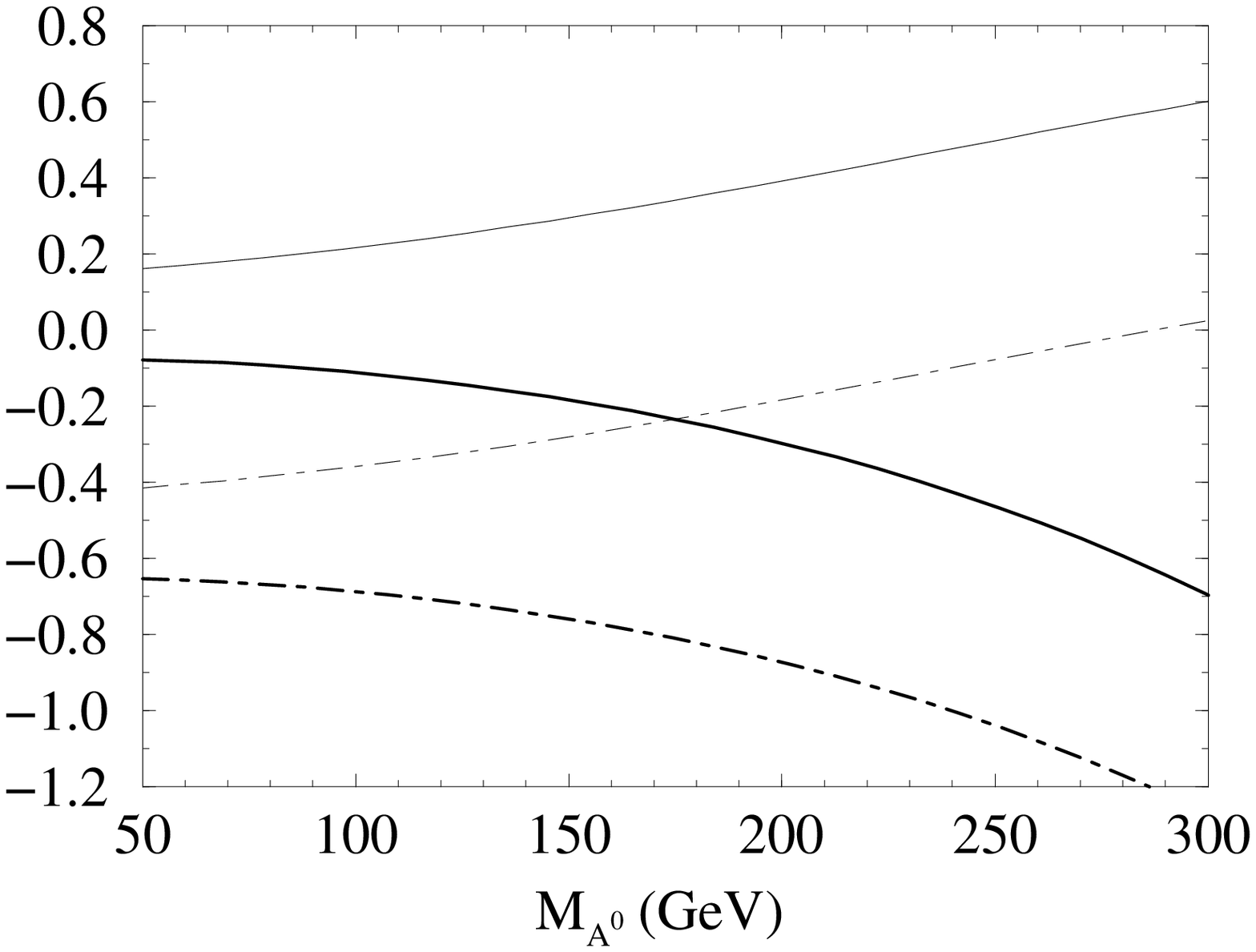,width=9.5cm} \\{\large (c)}\\\vspace{.2cm}\\
 \epsfig{file=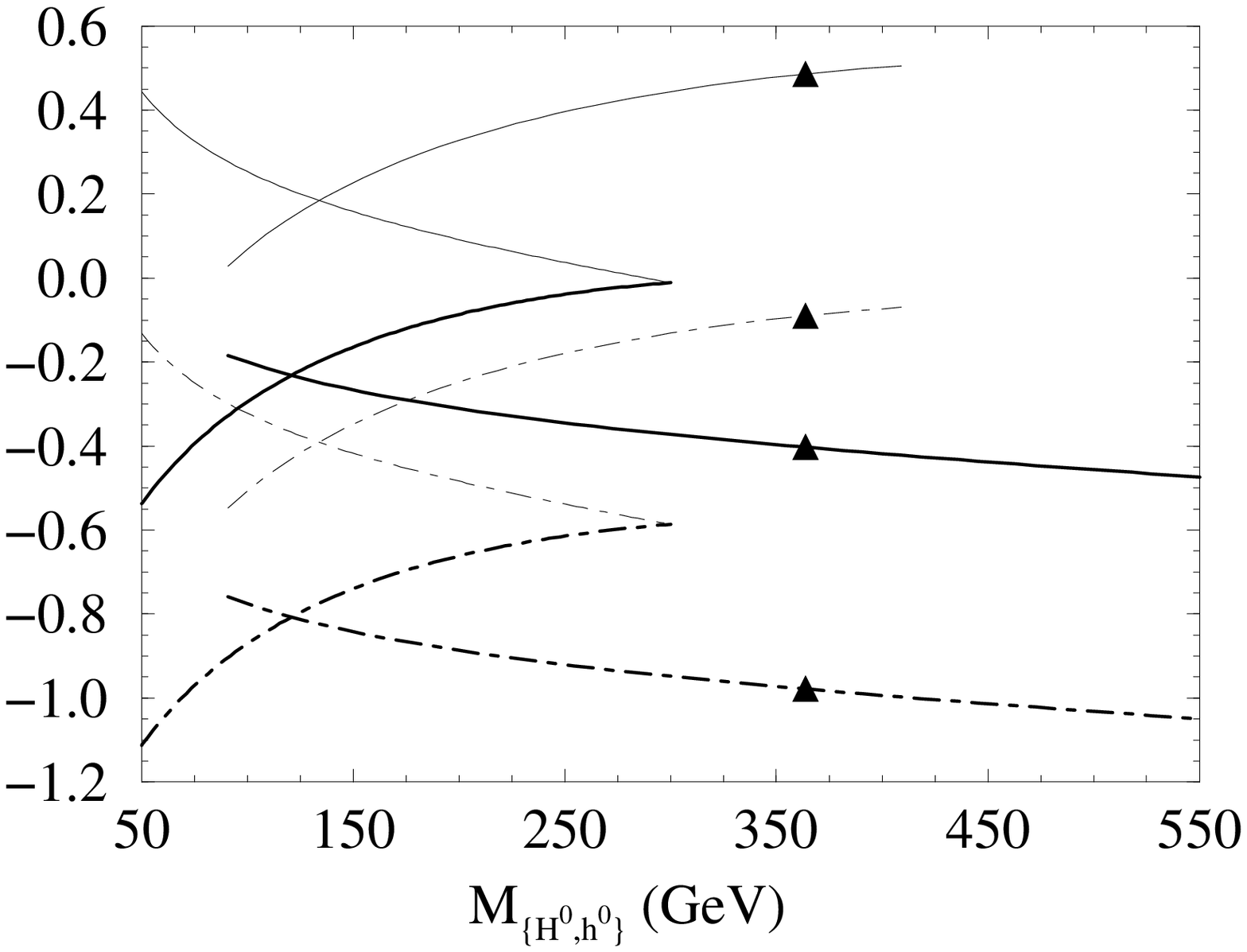,width=9.5cm} \\ {\large (d)} \\\vspace{.2cm}\\
{\Large Fig. 3 (cont.)}

\end{tabular}

\newpage

\begin{tabular}{cc}
\epsfig{file=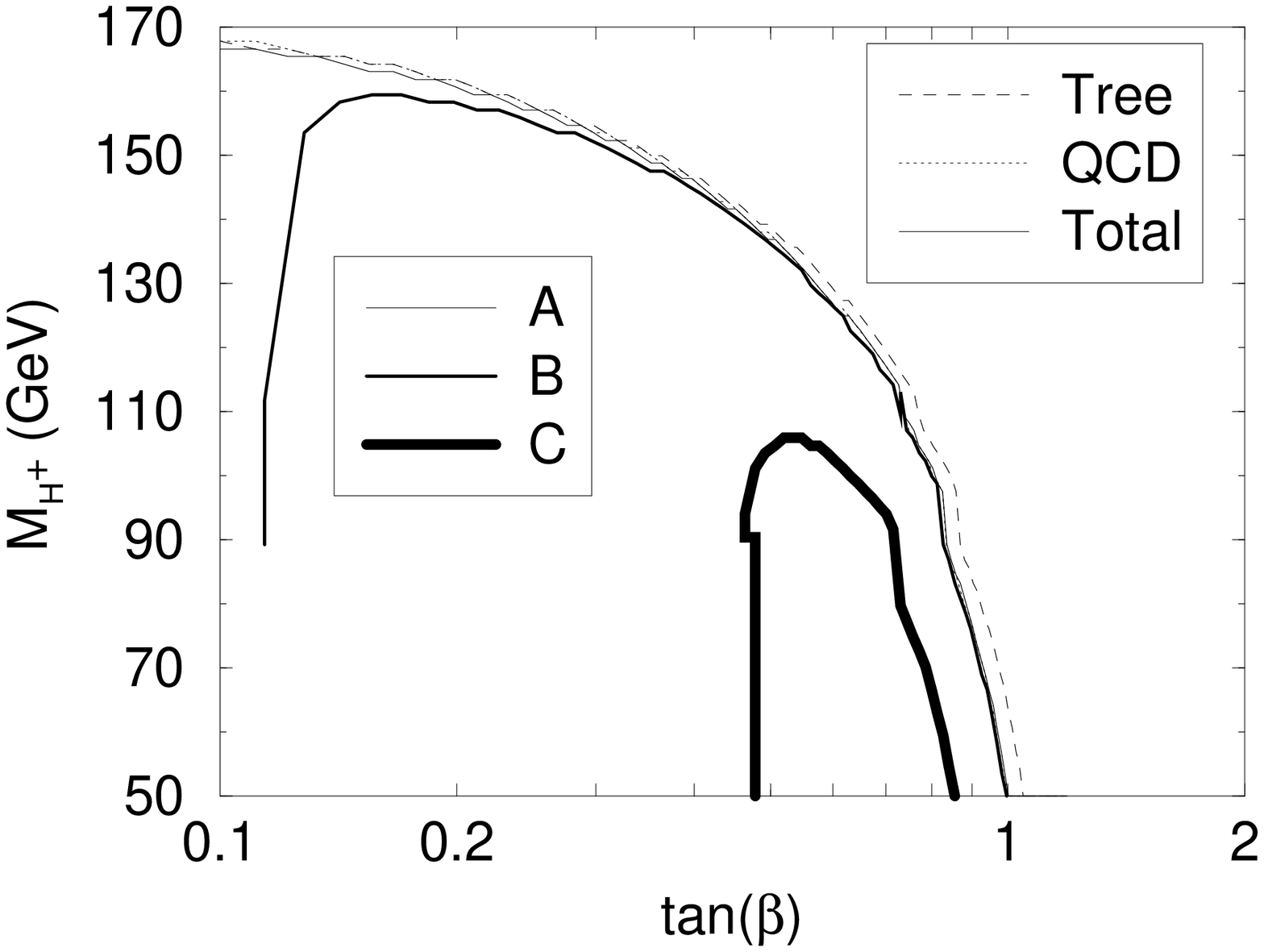,width=9.5cm} \\{\large (a)} \\\vspace{.2cm}\\
 \epsfig{file=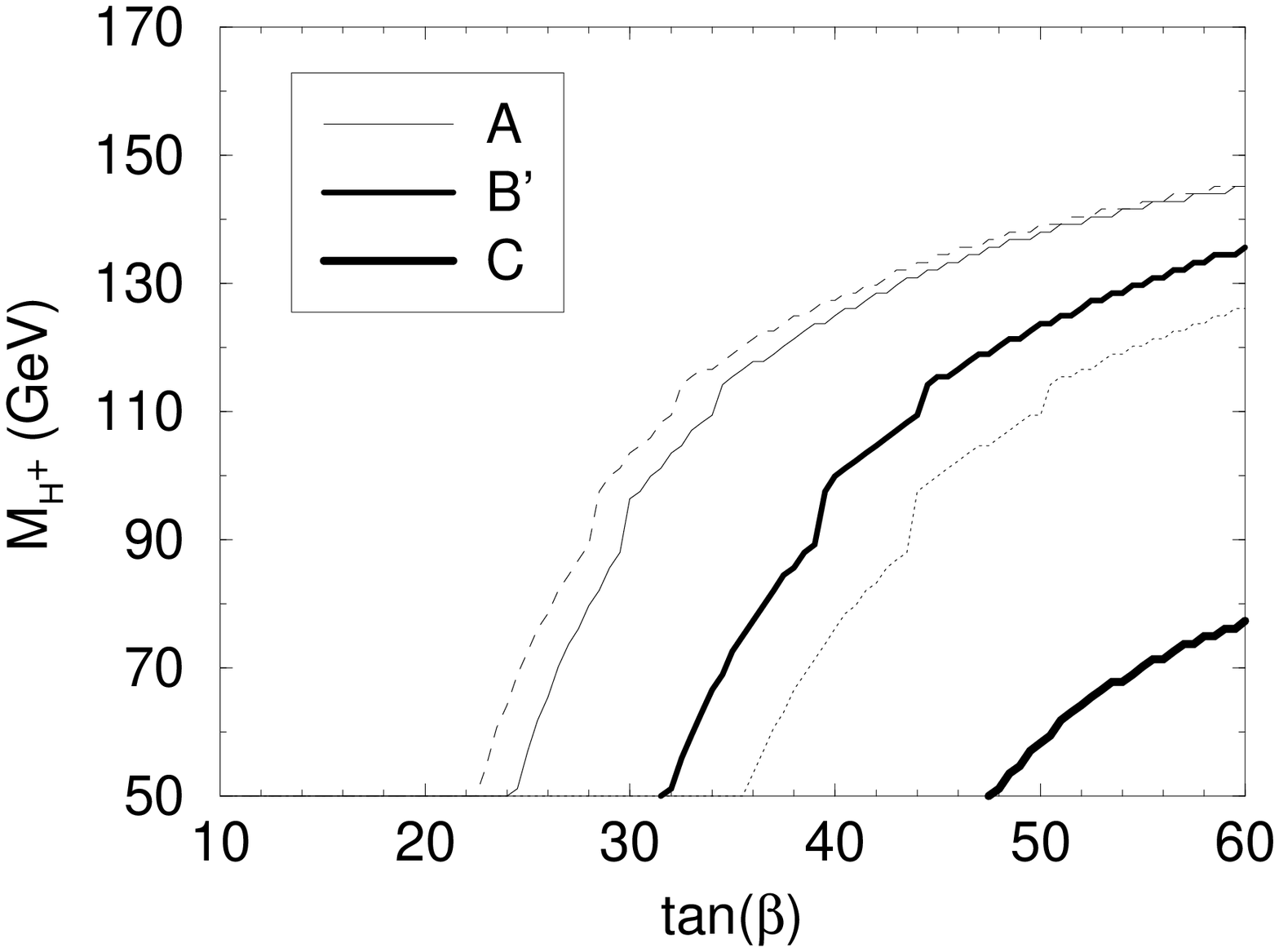,width=9.5cm} \\ {\large (b)} \\\vspace{.2cm}\\
{\Large Fig. 4}
\end{tabular}

\end{center}

\end{document}